\documentclass[preprint2]{aastex}

\shorttitle{The helium content of globular clusters}
\shortauthors{Villanova et al.}

\begin{document}


\title{The helium content of globular clusters: NGC6121 (M4)
   \thanks{Based on observations made with ESO telescopes at La Silla Paranal
   Observatory under programme ID 083.B-0083}.}


\author{S. Villanova and D. Geisler}
\affil{Departamento de Astronomia, Casilla 160, Universidad de Concepcion, Chile}
\email{svillanova,dgeisler@astro-udec.cl}

\author{G. Piotto}
\affil{Dipartimento di Astronomia, Universit\`a di Padova, Vicolo 
dell'Osservatorio 3, I-35122 Padua, Italy}
\email{giampaolo.piotto@unipd.it}

\and

\author{R.G. Gratton}
\affil{Osservatorio Astronomico di Padova, Vicolo dell'Osservatorio 
5, 35122 Padova, Italy}
\email{raffaele.gratton@oapd.inaf.it}




\begin{abstract}
In the context of the multiple stellar population scenario in globular clusters (GC), helium 
(He) has been proposed as a key element to interpret the observed multiple main sequences (MS), subgiant 
branches (SGB) and red giant branches (RGB), as well as the complex horizontal branch (HB) morphology.
In particular, second generation stars belonging to the bluer part of the HB, are thought to
be more He rich ($\Delta$Y=0.03 or more) but also more Na-rich/O-poor than those located in the redder part
that should have Y equal to the cosmological value.     
Up to now this hypothesis was only partially confirmed in NGC~6752, where
stars of the redder zero-age HB showed a He content of Y=0.25$\pm$0.01, fully
compatible with the primordial He content of the Universe, and were all Na-poor/O-rich.
Here we study hot blue horizontal branch (BHB) stars in the GC NGC~6121 (M4)
to measure their He plus O/Na content. Our goal is to complete the partial results obtained for
NGC~6752, focusing our attention on targets located on the bluer part of the HB of M4.
We observed 6 BHB stars using the UVES@VLT2 spectroscopic facility.
Spectra of S/N$\sim$150 were obtained and the very weak He line at 
5875 \AA\ measured for all our targets. We compared this feature with synthetic
spectra to obtain He abundances. In addition O, Na, and Fe abundances were estimated.
Stars turned out to be all Na-rich and O-poor and to have a homogeneous He content with a mean value of 
Y=0.29$\pm$0.01(random)$\pm$0.01(systematic), which is enhanced by $\Delta$Y$\sim$0.04 with respect
to the most recent measurements of the primordial He content
of the Universe (Y$\sim$0.24$\div$0.25). 
The high He content of blue HB stars in M4 is also confirmed by the fact that
they are brighter than red HB stars (RHB). Theoretical models suggest
the BHB stars are He-enhanced by $\Delta$(Y)=0.02$\div$0.03 with respect  to
the RHB stars.
The whole sample of stars has a metallicity of [Fe/H]=-1.06$\pm$0.02 (internal
error), in agreement with other studies available in the literature.
This is a rare direct measurement of the (primordial) He abundance for
stars belonging to the Na-rich/O-poor population of GC stars in a temperature regime 
where the He content is not altered by sedimentation or extreme
mixing as suggested for the hottest, late helium flash HB stars.
Our results support theoretical predictions that the Na-rich/O-poor
population is also more He-rich than the Na-poor/O-rich generation
and that a leading contender for the 2$^{nd}$ parameter is the He abundance.
\end{abstract}



\section{Introduction}

In the last few years, following the discovery of multiple sequences in the 
color-magnitude diagrams (CMD) of many globular clusters (GCs), the debate on their
He content has been renewed. In this respect, the most interesting and peculiar 
clusters are  $\omega$ Centauri and NGC~2808, where at least 3 main sequences 
(MS) are present \citep{Be04, Vi07, Pi07}. The color of these sequences cannot be 
explained in terms of a metallicity effect. In $\omega$ Centauri \citet{Pi05}
showed that the bluest MS is more metal rich than the main red population, 
while in NGC~2808 they all have the same iron content (as inferred from abundances 
in RGB stars). The only remaining parameter affecting significantly the position 
of a star in the MS is the helium content, and this was the explanation proposed 
for the photometric and spectroscopic properties of the MS stars in 
both clusters \citep{No04, Pi05, Da05, Da07, Pi07}, with the bluer MS
stars being more He enriched.

\begin{figure}[ht]
\epsscale{1.0}
\plotone{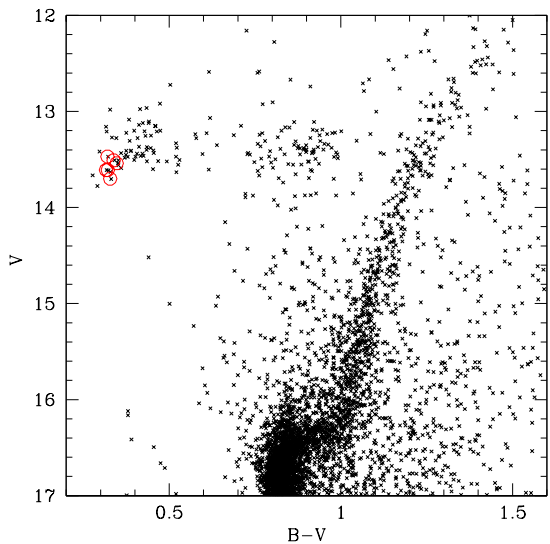}
\caption{The CMD of M4 with the observed BHB stars indicated as open circles.}
\label{f1}
\end{figure}

On the other hand almost all GCs {\bf observed to date} \citep{Ca09} show some kind of spread in their element
content at the level of the RGB, the most evident being the spread in Na and O, elements that are
anticorrelated \citep{Ca10}. Na and O abundances are also (anti)correlated with other light
elements, such as C,N,Mg, and Al \citep{Gr04}.

The most natural explanation for this phenomenon is that 
clusters experienced an extended  period of star formation, where the younger
populations were born from an interstellar medium polluted by products of the
CNO {\bf (and possibly NeNa, MgAl)}
cycle coming from massive stars of the former generation \citep[ and reference
therein]{Ca11}. In this picture the
interstellar medium is affected by an enhancement of its He content, together
with N, Na, and Al, while C and O turn out to be depleted.
This hypothesis can also explain correlations or anticorrelations 
of light elements present at the level of unevolved stars \citep{Gr01}.
This phenomenon cannot be due to evolutionary effects like
deep mixing processes that happen on the red giant branch (RGB) 
only after the first dredge-up (i.e. in stars brighter than
the RGB-bump), since it is also present in MS stars.

Evidence for a direct correlation between the He and Na abundances 
is  now accumulating. \citet{Br10a} found differences in the
abundances of Na as well as of other elements along the different
main sequences of NGC~2808. In another paper \citet{Gr10},
the same group found correlations between several expected features 
likely related to altered He abundances and the Na abundances for stars 
along the RGB for several clusters (the most clear evidence being again
for NGC~2808). Very recently, \citet{Du10} and \citet{Pa11}
observed that the chromospheric  He I 10830~\AA\ line is stronger in
Na-rich RGB stars in $\omega$~Cen and NGC~2808, respectively, than in Na-poor
stars.

\citet{Ca07} found that the extension of the Na-O anticorrelation 
is related to the extension of the horizontal branch (HB), suggesting that 
the anticorrelation may be related also to HB morphology.

Long ago \citet{No81}, measuring the
strength of the CN band at 3839 \AA\ in M4, was the first to speculate
on the possible connection between light-element and HB morphology
in M4. \citet{Ca95} were the first to speculate on
the connection between super-oxygen-poor stars and blue HB morphology
in M3 and M13.

It has been clear since the 1960s that the HB morphology is
related not only to the cluster iron content (as expected from the
models), but also to other parameters
(the so called {\it second parameter problem}) which must account 
for the fact that some clusters have an HB extended or extremely 
extended to the blue, while others of the same metallicity do not.

{\bf \citet{Gr10} showed that the main candidate to be the second parameter is age.
However they indicate also that at least a third parameter is required, and
that it is most likely He.}

A spread in He abundance can reproduce the HB morphology in GCs (as
first noticed by \citealt{Ro73}) (see also the extensive analysis by
\citealt{Gr10}). According to this picture, both stars with normal 
and enhanced He content end up on the zero-age horizontal branch (ZAHB) 
after the onset of core He burning. However, if mass loss in the previous
phase is similar, He-rich stars should be less massive (and thus warmer)
than He-normal ones because they burn H faster when they are on the main 
sequence. So He enhancement could provide the mechanism required to spread
stars from the redder and cooler HB (stars with normal composition) to the
hotter and bluer part of the HB (He-rich stars) as discussed in \citet{Da02}.

\begin{figure}[ht]
\epsscale{1.00}
\plotone{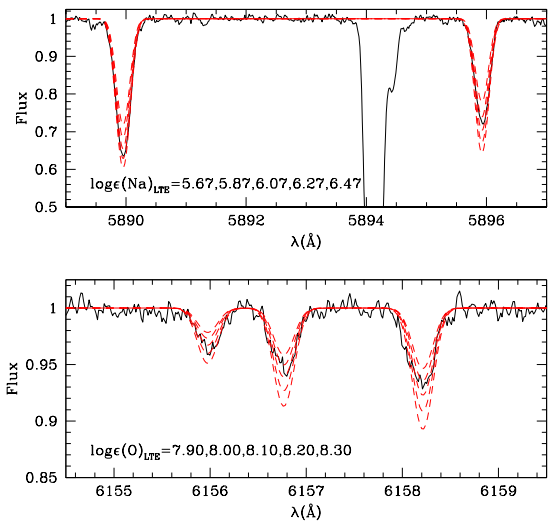}
\caption{Example of Na (upper panel) and O (lower panel) spectral line fitting
  for the star \#45025. Values for Na and O abundances of the synthetic
  spectra are indicated.}
\label{f2}
\end{figure}

However, the most direct evidence, the spectroscopic determination of He
abundances directly for the HB stars, is still scarce.
\citet[][hereafter Vi09]{Vi09} analyzed a sample of stars in NGC~6752 belonging
to the HB in the  T$_{\rm eff}$ range between
8000 and 9000 K. As discussed in that paper only stars between 8500 and 11500
K are good targets for this purpose because they  are sufficiently
hot to show features of He in their spectra but cooler than the Grundahl-jump \citep{Gr99},
the temperature limit above which stars show evidence of metal levitation and He
sedimentation, which alters the original surface abundances \citep{Pa06}.
Vi09 found for all ZAHB stars of the redder part of the blue HB in NGC~6752
a  Na-O content that, when compared with the Na-O anticorrelation found for
the RGB \citep{Ca09}, demonstrates that they belong to the Na-poor and O-rich 
(primordial) population. The only evolved HB target, which likely comes from
the bluer HB, belongs to the Na-rich, O-poor population.
But, most important, they show that stars of the redder ZAHB are all
Helium-normal (Y=0.25$\pm$0.01, where Y is the fractional mass of He),
in agreement with the proposed scenario.

In addition, very recently \citep{Ma11} found a direct
correlation between Na and O abundances and colours of the stars along the
HB of M4. In fact, RHB stars are all Na-poor and O-rich, while BHB are all
Na-rich and O-poor. According to this result we expect the BHB to be
He-rich, and the RHB to be He-normal.

While all of this data point toward a connection between He spread and HB
morphology, a key piece of information 
is still missing to finally confirm empirically this scenario.
We need to verify whether ZAHB stars suitable for He
measurement (8500$<$T$_{\rm eff}<$11500 K) and belonging to
the bluer HB indeed show a He-enhancement.
The aim of this paper is to fill this gap by measuring 
the He content in blue HB (hereafter BHB) stars of M4 (NGC~6121).
This cluster has been studied in detail \citep{Ma08}. It has a bimodal HB
with the hotter stars at T$_{\rm eff}\sim$9500 K, 
and a bimodal Na-O anticorrelation, which is possibly related to a spread in He abundance. 
For this reason it is the ideal target for our purposes. We want to
verify whether the hot HB stars of M4 are He-rich, as well as Na-rich/O-poor.
In Section 2 we describe the observations. In Sec. 3 we discuss the
determination of the atmospheric parameters of our stars and
the line list we used. In Sec. 4, 5, 6, 7, 8, and 9 we present the
spectroscopic and photometric analysis, compare our findings with the literature,
and discuss and summarize our results.

\begin{figure}[ht]
\epsscale{1.0}
\plotone{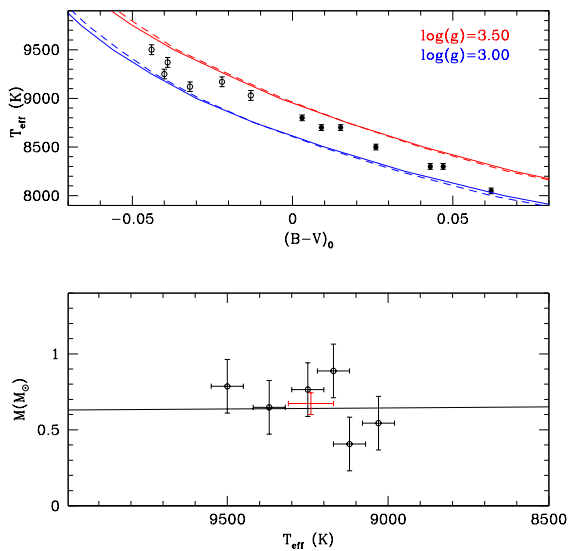}
\caption{T$_{\rm eff}$ as a function of the B-V color (upper panel)
         and mass as a function of T$_{\rm {eff}}$ (lower panel) for our sample of stars (open circles).
         Filled circles are stars of NGC~6752 from Vi09 for comparison.
         In the upper panel our data are compared with Kurucz's synthetic 
         colors for different gravities. Continuous lines are  
         synthetic colors for M4 metallicity, while dashed lines
         are synthetic colors for NGC~6752 metallicity.
         In the lower panel our data are compared with the theoretical mass
         (continuous line) for HB stars. The red cross is the mean value for
         our targets.}
\label{f3}
\end{figure}

\section{Observations, data reduction and membership}

Observations of stars in the field of view of M4
were carried out in June-October 2009
in the context of the ESO Program ID 083.B-0083.
In this program we observed a sample of 6 BHB stars (see Fig.\ref{f1})
for a total of 10$\times$45 min. exposures, selected from B,V photometry obtained
by the WFI imager at the ESO2.2m telescope.\\
The selected targets have spectral type A0 ( ${\rm (B-V) _0\sim0.0}$, T${\rm_{eff}\sim9000}$ K ).
This choice allows us to have targets showing He features in their spectrum,
but not affected by levitation or sedimentation.
Observations were performed with the FLAMES-UVES spectrograph
at the VLT@UT2(Kueyen) telescope. 
Spectra of the candidate BHB stars 
were obtained using the 580nm setting
with 1.0'' fibers, and cover the wavelength range 
4800-6800 \AA\ with a mean resolution of R=47000.\\
Data were reduced using the UVES pipeline \citep{Ba00},
where raw data were bias-subtracted, flat-field corrected,
extracted and wavelength calibrated.
Echelle orders were flux calibrated using the 
master response curve of the instrument. 
Finally orders were merged to obtain a 1D spectrum
and the spectra of each star sky-subtracted and averaged. Each spectrum has
a mean S/N$\sim$150 per resolution element at 5875 \AA.\\
The membership was established by radial velocity measurement.
We used the {\it fxcor} IRAF utility to measure radial velocities. This routine
cross-correlates the observed spectrum with a template having known radial
velocity. 
As a template we used a synthetic spectrum calculated for a typical A0III star 
(T$_{\rm {eff}}$=9000, log(g)=3.00, v$_{\rm t}$=1.00 km/s, [Fe/H]=-1.5, roughly
the same parameters as our stars).
Spectra were calculated using the 2.76 version of SPECTRUM,  the local
thermodynamical equilibrium spectral synthesis program freely distributed by
Richard O. Gray\footnote{See
http://www.phys.appstate.edu/spectrum/spectrum.html for more details.}.\\
The error in radial velocity - derived from {\it fxcor} routine - is less
than 1 km/s. Finally, for the abundance analysis, each spectrum was shifted
to rest-frame velocity and continuum-normalized.\\
Table~\ref{t1} lists the basic parameters of the selected stars:
the ID, the J2000.0 coordinates (RA \& DEC), V magnitude, B-V and U-V colors
\citep{Mo03}, heliocentric radial velocity (RV$_{\rm H}$),
T$_{\rm {eff}}$, log(g), microturbulence
velocity (v$_{\rm t}$, for determination of atmospheric
parameters see Sections 3 and 4).
From the measured radial velocities we obtained a mean heliocentric radial velocity
and a dispersion of:

\begin{center}
${\rm <RV_H>=70.2\pm2.6\ km/s},\ {\rm \sigma_{RV}=6.8\pm1.8\ km/s}$
\end{center}

The mean velocity agrees well with literature values ( see e.g. \citet{So09}:
${\rm <RV_H>=70.3\ km/s}$ ).
The dispersion we found is  quite high, but also its error is large.
It agrees within 2$\sigma$ with the more recent value
derived in the literature \citep[3.5$\pm$0.3 km/s]{Pe95}.
Considering the fact that at the position on the CMD of  the target stars
there is no significant background
contamination, and that their [Fe/H] content agrees very well with the mean
value for the cluster (see Section 4), we conclude that all the observed stars are cluster members.

\begin{deluxetable}{lcccccccccc}
\tablecolumns{10}
\tabletypesize{\scriptsize}
\tablecaption{Basic parameters of the observed stars.}
\tablewidth{0pc}
\tablehead{
\colhead{\tiny ID} & \colhead{\tiny RA(${\rm hh:mm:ss}$)} & \colhead{\tiny DEC(${\rm^o:\arcmin:\arcsec}$)} & 
\colhead{\tiny V(mag)} & \colhead{\tiny B-V(mag)} & \colhead{\tiny U-V(mag)} & \colhead{\tiny RV$_{\rm H}$(km/s)} & 
\colhead{\tiny T$_{\rm {eff}}$(K)} & \colhead{\tiny log(g)(dex)} & \colhead{\tiny v$_{\rm t}$(km/s)} & \colhead{\tiny vsini (km/s)}
}
\startdata
45025 & 16:23:34.380 & -26:32:36.60 & 13.54 & 0.35 & 0.85 & 80.864 & 9030 & 3.30 & 1.57 &  9\\  
46061 & 16:23:47.820 & -26:32:06.00 & 13.47 & 0.32 & 0.77 & 75.784 & 9250 & 3.45 & 1.42 &  3\\
47570 & 16:23:34.760 & -26:31:24.70 & 13.61 & 0.32 & 0.77 & 66.577 & 9370 & 3.45 & 1.02 &  7\\ 
49034 & 16:23:37.080 & -26:30:44.60 & 13.61 & 0.31 & 0.73 & 62.305 & 9500 & 3.55 & 0.90 &  7\\ 
49412 & 16:23:36.300 & -26:30:34.00 & 13.51 & 0.34 & 0.84 & 74.495 & 9170 & 3.52 & 1.50 &  5\\ 
50996 & 16:23:27.760 & -26:29:49.00 & 13.70 & 0.33 & 0.78 & 66.677 & 9120 & 3.25 & 0.95 & 10\\ 
\enddata
\label{t1}
\end{deluxetable}

\section{Atmospheric parameters, rotation and chemical abundances}

We used the abundances from FeI/II
features to obtain atmospheric parameters using the equivalent width (EQW)
method.\\ 
None of our stars show evidence for {\bf strong} rotation {\bf (see Table~\ref{t1})}.
For this reason EQWs are obtained from a Gaussian fit
to the spectral features.\\
We could measure only a small number of Fe lines for each spectrum (5
FeI lines and 9 FeII lines) due to the limited spectral coverage. 
However the high quality of our spectra (allowing an
accurate measurement of the EQWs) and the simultaneous use of both FeI/II sets
of lines allowed us to obtain reliable atmospheric parameters, as confirmed by
the comparison of theoretical and observational results (see below).
The analysis was performed using the 2009 version of MOOG \citep{Sn73}
under a LTE approximation coupled with ATLAS9 atmosphere models \citep{Ku92}.\\

T$_{\rm eff}$ was obtained by eliminating any trend in the relation of 
the abundances obtained from Fe~I and Fe~II lines with respect to the
excitation potential (E.P.), while  microturbulence velocity 
was obtained by eliminating any 
slope of the abundances obtained from FeI and FeII lines vs. reduced EQWs.
log(g) values were estimated from the ionization equilibrium of FeI and FeII
lines in order to have:

\begin{center}
${\rm log\epsilon(FeI)=log\epsilon(FeII)}$
\end{center}

where $log\epsilon(El.)$=$log(N_{El.}/N_{H})+12$. N$_{El.}$ and N$_{H}$
are the density of the element and of hydrogen in number of particles per cm$^3$.
Adopted values for the atmospheric parameters are reported in
Table~\ref{t1}.\\
All the targets, according to their position on the CMD, are consistent with
being ZAHB objects (see Fig.~\ref{f1}).\\
The typical random error in T$_{\rm eff}$ and v$_{\rm t}$ can be obtained by the following
procedure. First we calculated, for each star, the errors associated with the
slopes of the best least squares fit in the relations between abundance
vs. E.P. and reduced EQW. The average of the errors corresponds to the typical error on the
slopes. Then, we selected one star representative of the entire sample
(\#46061). We fixed the other parameters and varied first
temperature and then microturbulence until the slope of the line that best fits the relation between
abundances and E.P. or reduced EQW becomes equal to the respective mean
error. The amount of temperature and microturbulence variation represent an estimate of the random errors,
that turned out to be 50 K and 0.04 km/s respectively.
The error in gravity was estimated by satisfying the following equation:

\begin{center}
${\rm {\tiny log\epsilon(FeI)-\Delta log\epsilon(FeI)=log\epsilon(FeII)+\Delta log\epsilon(FeII)}}$
\end{center}
\noindent
where ${\rm \Delta log\epsilon(FeI)}$ and ${\rm \Delta log\epsilon(FeII)}$ are
the errors on FeI/II abundance as given by MOOG. 
In other words we took the values ${\rm log\epsilon(FeI)}$
and ${\rm log\epsilon(FeII)}$ that satisfied the ionization equilibrium, decreased
${\rm log\epsilon(FeI)}$  and increased ${\rm log\epsilon(FeII)}$ 
by the errors given by MOOG, and estimated a new gravity using these new
FeI/II abundances. The difference with the previous value was assumed
to be our error on gravity, that turned out the be 0.06 dex on average.
These errors are to be considered as random and internal. Systematic ones were
checked as described in the next Section.

For a detailed description of the linelist we used for He, O, Na, and Fe and
the solar value we adopted see Vi09. Suffice to say here that O and Na abundances were obtained by 
comparison with synthetic spectra from the O triplet at 6156-6158 \AA\ and the Na doublet at
5889-5895 \AA\ (see Fig.~\ref{f2} for an example).
Na and O are known to be affected by NLTE. For this reason, we applied the corrections by \citet{Ta97} and
\citet{Ma00}, interpolated to the atmospheric parameters of our stars.
On the other hand, as shown by Vi09, He abundances are not affected by NLTE,
probably because the He line at 5875 \AA, due to the very high E.P.,
is formed in very deep layers of the atmosphere, where departure from the LTE
condition is negligible due to the high density of the gas.
A further discussion is required about NLTE correction for Fe. Some authors
(i.e. \citealt{Qi01}) claim that for A type stars like Vega or our targets a
NLTE correction of +0.3 dex must be applied to FeI LTE abundances, while FeII
are formed in LTE approximation. Slightly smaller non-LTE corrections
have been recently estimated for such stars by \citet{Mas11}.
All our tests show instead that no NLTE
correction is required for FeI, at least down to log(g)=3.0. First of all,
the analysis for Vega done in Vi09 assuming LTE gave us the same abundance for
FeI and FeII within 0.02 dex, comparable with the r.m.s\ scatter.
This result was confirmed by the analysis of the
other targets of Vi09 and by \citet{Vi09b}. In particular gravities of Vi09 were obtained
from the wings of H Balmer lines, which are formed in LTE approximation.
In spite of that no appreciable difference was found in the mean FeI and FeII
abundances of the stars ($\Delta$[Fe/H]=0.04$\pm$0.05 dex).

There is a further effect to take into account. Our results indicate that
all the stars are He-rich (Y$\sim$0.29, see next section). In spite of that for our analysis
we used atmosphere models with normal He-content (Y$\sim$0.25).
The question is whether this difference has some impact on the final abundances.
We answered this by calculating with ATLAS9 a new He-enhanced atmosphere model
for the star \#46061, considered as representative of our sample.
Then we recalculated the abundances. We found that
O, Na, and Fe do not change in a significant way (0.01 dex or less).
On the other hand, the He content changes by $\Delta$log$\epsilon$(He)=+0.03 dex. 
The reason could be that He lines form deep in the atmosphere where temperature is higher
and the UV flux, strongly dominated by H and He opacity, is greater. 
This changes the structure of the atmosphere in the
deepest layers and the strength of the spectral lines that are formed there.
We took into account this effect by applying a correction
$\Delta$log$\epsilon$(He)=+0.03 dex  {\bf ($\Delta$Y=0.015)}
to the values for He obtained assuming He-normal atmosphere models.

{\bf Using spectral synthesis we could also measure projected rotational velocities for
each star. For this purpose we assumed a combined instrumental+rotational
profile for spectral features. The instrumental profile was assumed to be
Gaussian with FWHM=R/$\lambda$ (where R is the resolution of the
instrument). Then vsini was varied in order to match the observed profile of
Na-double lines. Results are reported in Table~\ref{t1}. The typical error on
vsini is 1-2 km/s \citep{Vi09}. The results confirm all these stars are
relatively slow rotators}

\begin{deluxetable}{lcccccc}
\tablecolumns{7}
\tabletypesize{\scriptsize}
\tablecaption{Abundances obtained for the target stars.}
\tablewidth{0pc}
\tablehead{
\colhead{El.} & \colhead{45025}  & \colhead{46061} & \colhead{47570}  & 
\colhead{49034}  & \colhead{49412}  & \colhead{50996}
}
\startdata
${\rm log\epsilon(He)}$ & 11.00  & 10.95 & 11.02  & 11.07  & 10.94  & 11.08\\
${\rm Y}$               &  0.28  &  0.26 &  0.30  &  0.32  &  0.26  &  0.32\\
${\rm [O/Fe]}$          &  0.34  &  0.36 &  0.30  &  0.34  &  0.27  &  0.38\\
${\rm [O/Fe]_{NLTE}}$   &  0.23  &  0.25 &  0.18  &  0.21  &  0.16  &  0.26\\
${\rm [Na/Fe]}$         &  0.82  &  0.85 &  0.69  &  0.66  &  0.73  &  0.61\\
${\rm [Na/Fe]_{NLTE}}$  &  0.47  &  0.55 &  0.39  &  0.37  &  0.40  &  0.28\\
${\rm [Fe/H]}$          & -1.07  & -1.06 & -1.04  & -1.04  & -0.99  & -1.13\\
\enddata
\label{t2}
\end{deluxetable}

\begin{figure}[ht]
\epsscale{1.00}
\plotone{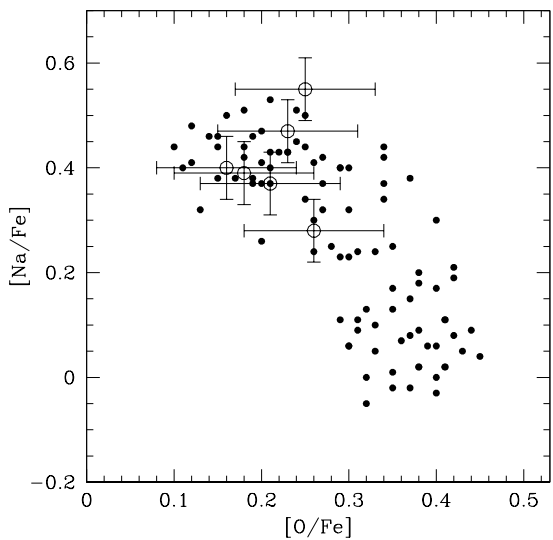}
\caption{Na-O abundances found for our HB targets (open circles). 
         Filled circles are data for RGB stars from Marino et al. (2008).
         See Sec. 4 for more details.}
\label{f4}
\end{figure}

\section{Systematic errors}
An estimation of the systematic errors  (or at least upper limits) is very
important for our purpose to compare our He abundance with the primordial
value of the Universe.

First of all we checked our T$_{\rm eff}$ scale comparing the observed colors
with synthetic B,V photometry, both from Kurucz\footnote{http://kurucz.harvard.edu/}.
For this purpose we assumed an E(B-V)=0.36 value from \citet{Ha96}. For Vi09 this
test was not satisfying because temperatures from colors were very different
from the spectroscopic ones. Meanwhile, we have further investigated this problem.
By comparing our photometry, based on observations taken with the wide-field imager at
the ESO2.2m telescope both for NGC~6752 and M4, with Stetson's database
\footnote{http://www3.cadc-ccda.hia-iha.nrc-cnrc.gc.ca/community/STETSON/standards/}
\citep{St00}, we found that the blue and red parts of our CMDs are affected by photometric
calibration problems, that can reach several hundredths of a magnitude.
As a consequence we corrected our photometry and the result is plotted in
Fig.~\ref{f3} (upper panel). Empty circles are M4 stars, while filled ones are
the old NGC~6752 data corrected for the reddening of the cluster.
T$_{\rm eff}$ and dereddened B-V colors were compared with Kurucz synthetic
photometry for log(g)=3.00 (blue) and 3.50 (red), which is roughly the gravity
interval our stars cover. Continuous lines have [Fe/H]=-1.00, while dashed
lines have [Fe/H]=-1.50 in order to cover the metal content of the two
clusters.
We note that colors in this temperature regime are very dependent on gravity
but are almost unaffected by metallicity.
The scatter for M4 stars is a bit higher than that for NGC~6752, due to the
differential reddening affecting the former cluster.

If we compare T$_{\rm eff}$ from B-V colors with those obtained from
spectroscopy for the two clusters, we find that for M4 the latter are
underestimated by about 100 K, while for NGC~6752 they are overestimated by about
200 K. We are left with a net difference of 100 K. Part or all of this
discrepancy is related to uncertainties in the value we adopted for the
interstellar reddening, because even a small error of 0.01 mag would lead to
an error of $\sim 150$~K in the effective temperatures. On average, the good
agreement between temperatures from colours and from spectra is then
comforting, and we can assume 150 K as un upper limit for the systematic
error in temperature.

A better indication concerning the T$_{\rm eff}$ systematic errors comes from the
comparison between our [Fe/H] (-1.06, see next section)
with \citet[-1.07]{Ma08} and \citet[-1.12]{Ma11}.
First of all we note that both datasets were taken with the same instrument, so
systematic effects on abundances due to the spectrograph (as happens for example
comparing abundances measured with UVES and GIRAFFE, \citealt{Ca09}) can be ruled out.
The difference is only 0.01 dex with respect to \citet{Ma08} and 0.06 dex
with respect to \citet{Ma11}. On the other hand, an error as large as 
100~K on our spectroscopic T$_{\rm eff}$\ would imply an offset of  
0.10~dex. We can see that in the worst case comparison with \citet{Ma11} gives
50 K as an estimate for our systematic error in temperature.

Finally \citet{Ma11} checked the reliability of their atmospheric parameters
comparing their T$_{\rm eff}$ with those derived from models by \citet{Da02}.
Their sample of stars distributes with a dispersion of $\sim$50 K and has a null
offset with respect to the line of perfect agreement (see their Fig. 2). This
implies a negligible systematic error on temparature, valid also for our data because
in the two papers targets are similar and the methodology is the same.

After these tests we assume $\Delta$T$_{\rm eff}$=50 K as the
systematic error on our T$_{\rm eff}$ scale but we will consider also the
upper limit $\Delta$T$_{\rm eff}$$<$150 K in the discussion.

Systematic errors on our gravity scale can arise from the fact that
we use FeI/II balance to obtain log(g). As said in the previous section
this can introduce a systematic shift if FeI lines are formed in 
NLTE (while FeII lines form in LTE), as suggested by some authors. Thanks to
our results on Vega and on other A stars presented in Vi09, we think we have
shown that FeI lines can be safely treated with LTE approximation, and as a
consequence systematic errors on the log(g) scale should be negligible. 

This statement is further supported by \citet{Yu11}.
Here the authors analize a T$_{\rm eff}$=12045 K, log(g)=3.9 dex star
using Kurucz's models, as in our case. They obtain atmospheric parameters from
Str\"{o}mgren photometry. As they say, {\it ATLAS9 model with the parameters
T$_{\rm eff}$=12045 K, log(g)=3.9 derived from the Str\"{o}mgren photometry
meets the requirement of same iron abundance from all the different kinds of
iron lines}. In fact, they obtain the same abundance for FeI and
FeII lines. If this is true for a 12000 K star, where NLTE effect (if
present) should be larger than in our colder stars due to the stronger iron
overionization, we can safely assume that LTE works well as an approximation for
the atmosphere models of our targets, and it can be used to obtain unbiased gravities.

However we decided to perform further tests.
For this purpose we calculated the mass ($\frac{M_{\star}}{M_{\odot}}$) 
of our stars by inverting the canonical equation:

\begin{center}
${\rm log(\frac{g_{\star}}{g_{\odot}})=4 \cdot log(\frac{T_{\star}}{T_{\odot}}) 
- log(\frac{L_{\star}}{L_{\odot}}) + log(\frac{M_{\star}}{M_{\odot}})}$
\end{center}
\noindent
where

\begin{center}
${\rm log(\frac{L_{\star}}{L_{\odot}})=-\frac{M_{bol}-4.74}{2.5}}$
\end{center}
\noindent
and

\begin{center}
${\rm M_{bol}=M_V+BC=V-(m-M)_V+BC}$
\end{center}

Bolometric correction (BC) was taken from \citet{Fl96}, and
distance modulus $(m-M)_V$ from \citet{Ha96}.
Results and relative errors are reported in Fig.~\ref{f3} (lower panel) and compared
with the theoretical model from \citet{Mb07}. The red cross is the mean value for our
stars. It agrees very well with the theoretical value, well within 1$\sigma$.
To match exactly the theoretical value we should change our log(g) scale by 0.03 dex.

After these tests we conclude that $\Delta$log(g)=0.05 dex is a reasonable
value as the systematic error for our gravity scale.

We finally can estimate an upper limit for the systematic error on the He
abundance. $\Delta$T$_{\rm eff}$=50 K gives $\Delta$log$\epsilon$(He)=0.02
dex, and $\Delta$log(g)=0.05 dex again gives $\Delta$log$\epsilon$(He)=0.02.
Those two values each translate into $\Delta$Y=0.01 and 0.01 respectively.
The squared sum gives:

\begin{center}
${\rm \Delta_{sys}(Y)=0.01\ (the\ exact\ value\ is\ 0.014)}$
\end{center}

which is our systematic error on the He abundance.
If we consider the upper limit $\Delta$T$_{\rm eff}$=150 K instead, we
end up with:

\begin{center}
${\rm \Delta_{sys}(Y)<0.03}$
\end{center}

Both values will be used in the discussion.

\section{Results of the spectroscopic analysis}

The chemical abundances we obtained are summarized in Table~\ref{t2}.
From our sample the cluster turns out to have a Fe content of:

\begin{center}
${\rm [Fe/H]=-1.06\pm0.02}$
\end{center}

(internal error only) which agrees very well with the results from \citet{Ma08} 
([Fe/H]=-1.07$\pm$0.01). The agreement is slightly worse but within 0.06 dex
 with respect to \citet{Ma11}{\bf ([Fe/H]=-1.12)}. 

In Fig.~\ref{f4} we compare Na and O abundances for our target stars
with the Na-O anticorrelation found by \citet{Ma08} for a sample of M4 RGB stars.
\citet{Ma08} identified two separated populations in the Na-O plane,
one O-rich and Na-poor, the other O-poor and Na-rich.
We find that all our blue HB stars have a Na/O content which is fully
compatible with the O-poor/Na-rich population. None of our targets belongs to
the O-rich/Na-poor group. 
This is an indication that the light-element
spread (Na and O in this case) is a vital clue to the morphology of the HB,
and suggests that all O-rich/Na-poor objects end-up on the redder part of the ZAHB, while O-poor/Na-rich
stars end-up on the bluer part of the ZAHB.
We show this statistically by calculating the probability
to find by chance six stars all belonging to the  O-poor/Na-rich
population, under the hypothesis that the HB morphology does not depend on the
O/Na chemical content. \citet{Ma08} show that the two populations in M4
contain about 50\% of the total stars each. Under the previous hypothesis,
we expect to find the blue HB to have an equal mixture of the two populations.
The probability of finding 6 stars, all O-poor/Na-rich, as we found
for the blue HB stars, is less than 2\%. 
Therefore, we can conclude, at 98\% confidence level, that the
HB position does depend on the O/Na chemical content, or on a related
factor, such as the He abundance.

A further confirmation of this assessment comes from \citet{Ma11}.
In that paper we analyzed a large sample of stars of the two HBs in M4.
For a subsample, observed with UVES, we could measure both O and Na, while
for the remaining stars, observed with GIRAFFE and belonging to the red HB,
only Na. We found that all the blue HB objects are
O-poor/Na-rich, as in the present paper. On the other hand the two red HB
stars with measured O are O-rich/Na-poor. 
All the remaining red HB stars for which we collected GIRAFFE spectra
have Na that is compatible 
with the O-rich/Na-poor population.
Again, the probability to find this result by chance is extremely low,
in this case well below 1\%.

The clear conclusion is that, at least for M4, the HB morphology is related
to {\bf the light element content, specifically Na and O}.

\begin{figure*}
\epsscale{2.50}
\plottwo{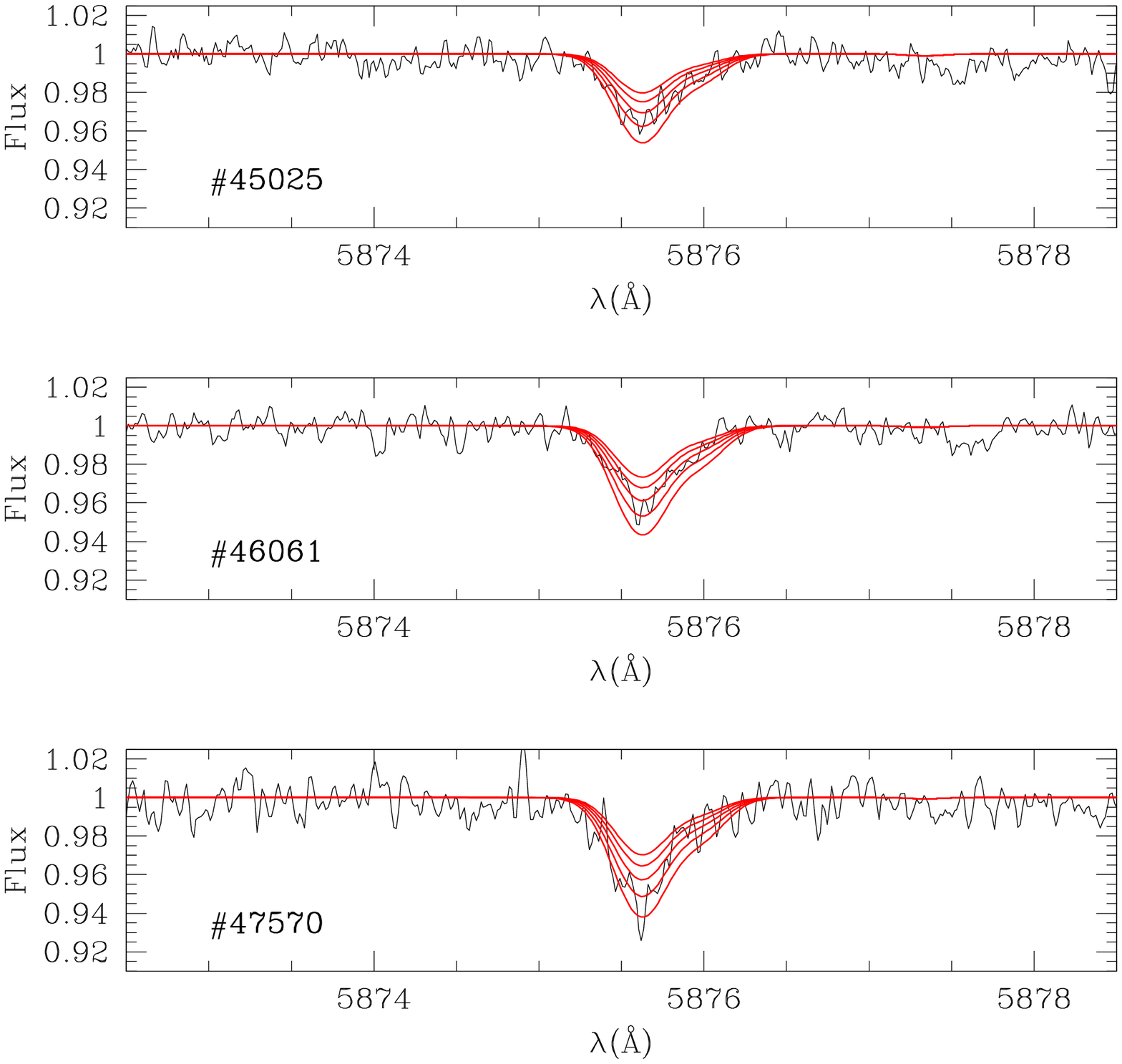}{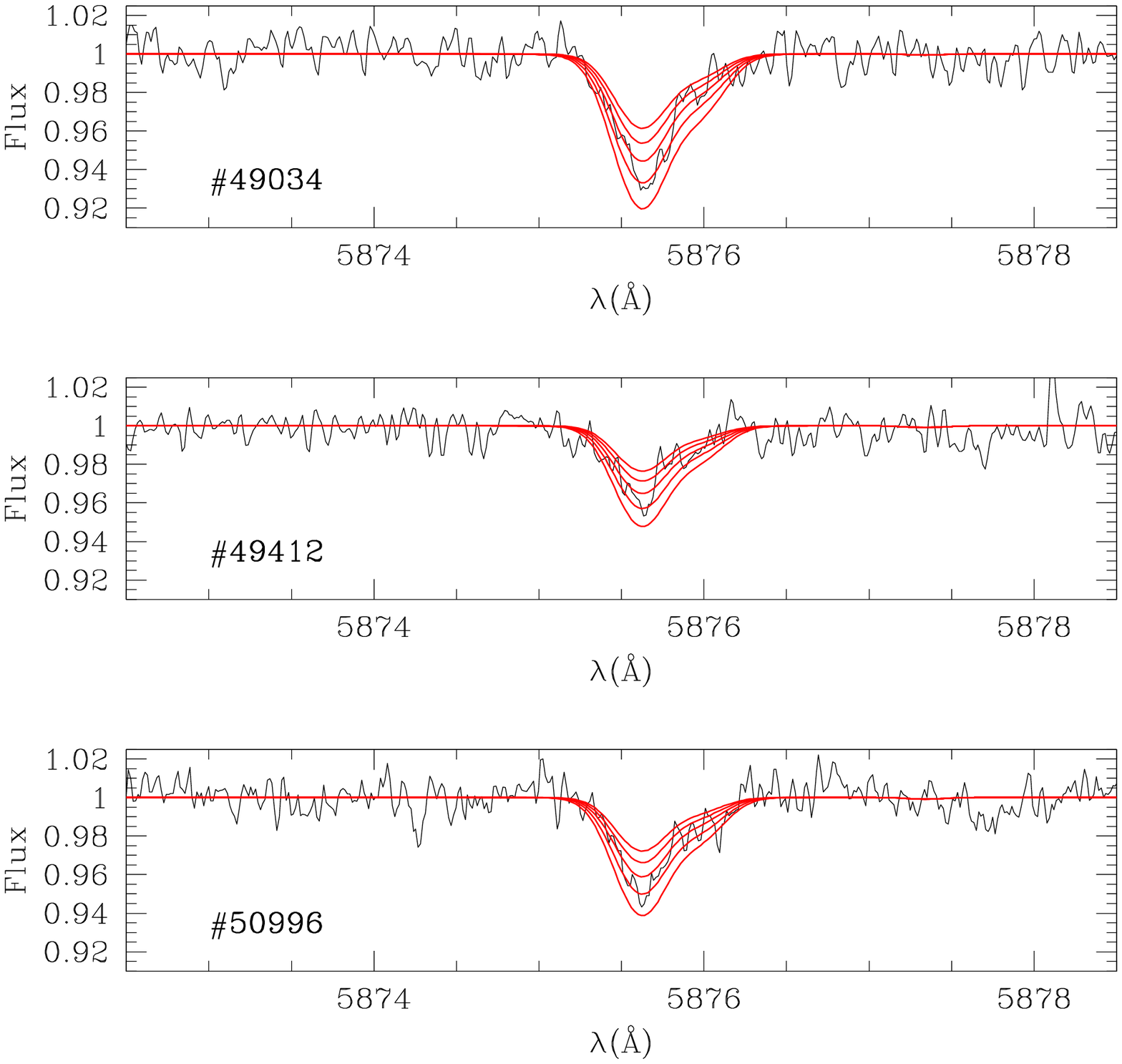}
\caption{He lines of our stars. We superimposed on each observed spectrum 
         5 synthetic ones with log$\epsilon$(He)=10.70,10.80,10.90,11.00,11.10
         respectively.}
\label{f5}
\end{figure*}

But a more important result regards the He content.
In Fig.~\ref{f5} we compare the observed spectra around the He line
with 5 synthetic spectra with different He content. The stars turn out to have
a mean He content of ${\rm log\epsilon (He)}$=11.01$\pm$0.02. This translates into:

\begin{center}
${\rm Y=0.29\pm 0.01}$
\end{center}

{\bf In Tab.~\ref{t2} logaritmic He abundances ${\rm log\epsilon (He)}$ for
single stars were trasformed in mass fraction Y value as well.}
As noted before, our best estimation of the systematic error on the He
measurement is:

\begin{center}
${\rm \Delta_{sys}(Y)=0.01\ (or\ 0.014)}$
\end{center}

while the upper limit is:

\begin{center}
${\rm \Delta_{sys}(Y)<0.03}$
\end{center}

We conclude that the He content of our stars is larger than that of the primordial 
Universe (Y$\sim$0.24$\div$0.25) with a level of confidence of about 4$\sigma$
if we only consider the internal error. If we combine internal and 
systematic errors using the squared sum, the level of confidence is lower but
still more than 2$\sigma$. It is less than 2$\sigma$ (between 1.3 and 1.6$\sigma$,
depending on if we assume Y=0.24 or Y=0.25 for the primordial content) only if we
use the upper limit for the systematic error.
A possible point against the significance of our result comes from
\citet{Sw87}. This paper suggests that after the first dredge-up, the Y content
of a star increases by $\sim$0.015 for Z=0.001 (roughly the metallicity of
M4). So we should compare our absolute abundance with 
Y$\sim$0.24$\div$0.25+0.015$\sim$0.26. This would lower our significance
to 1.7$\sigma$ (1.0$\sigma$ if we use the upper limit for the systematic
error). {\bf However this He enhancement due to the first dredge-up is
controversial, because other mechanisms (e.g. atomic diffusion, radiative
acceleration, and turbulence) could be at work and playing a
role in defining the precise He difference between MS and HB}. 

While not definitively proven, we believe our data strongly hints at a He
abundance larger than the primordial value or than the value expected for a
He-normal star after the first dredge-up. In the following, we will give
various arguments that definitely support a similar conclusion.

First, we can compare the present results with the He content of HB stars of NGC 6752 found by 
Vi09. In both studies, we used the same spectrograph and methodology to estimate He abundance.
Therefore, in such a comparison any systematic error is cancelled out.
Vi09 found Y=0.25$\pm$0.01 for their sample of stars all belonging to the redder part of
the blue HB of that cluster. Such a value is compatible with the primordial He abundance of
the Universe. The stars of the present study are instead located on the bluer part of the HB of M4,
and are then expected to be He-enhanced.
The Y values of the two samples differ at the level of 2.8$\sigma$.
We also applied a Kolmogorov-Smirnov test to the two datasets. This test 
concludes that the two distributions are not compatible, with
a confidence level of more than 90\%.
We note that in this case the \citet{Sw87} result does not affect the comparison because
BHB targets in both clusters should have experienced the same
surface He-enhancement after the first dredge-up.
{\it Therefore we conclude that M4 BHB stars are He-enhanced}.

\begin{figure*}[ht]
\epsscale{1.50}
\plotone{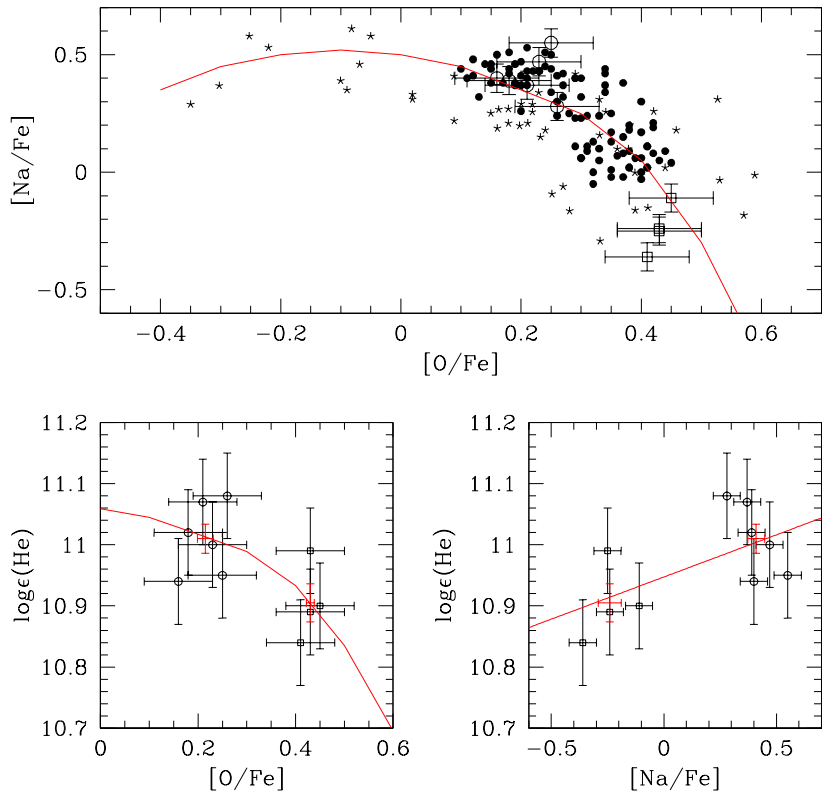}
\caption{Upper panel: Na-O anticorrelation as defined by RGB and HB stars of M4 and
  NGC~6752. Lower left panel: He-O anticorrelation as defined by HB stars of M4 and
  NGC~6752. Lower right panel: He-Na correlation as defined by HB stars of M4 and
  NGC~6752. See text for more details}
\label{f6}
\end{figure*}

We next compare M4 and NGC~6752 in more detail including O and Na.
For this purpose we plot in Fig.~\ref{f6} (upper panel) O and Na abundances
from the following sources: M4 RGB stars (filled circles) from \citet{Ma08},
NGC~6752 RGB stars (stars) from \citet{Ca07b}, M4 HB stars (open
circles, this work), and NGC~6752 HB stars (open squares) from \citet{Vi09}.
We see that RGB stars of the two clusters define a common Na-O anticorrelation
fitted by the curve. NGC~6752 stars cover a larger range both in O and Na.
HB stars follow the same curve, as expected, but they map only
a limited part of the anticorrelation, due to the fact that they are
located in a restricted part of the HB.
According to the results of this paper, the Na-O anticorrelation is accompanied
by a He-O  anticorrelation. This is shown in Fig.~\ref{f6} (lower left panel).
Red crosses are the mean values and errorbars for the two groups of stars, while
the curve is the fit to the Na-O trend shifted and compressed in the y
direction in order to match the observed points. This curve represents the He
content that a stars has according to its Na abundance. In Fig.~\ref{f6} (lower
right panel) we report also the He-Na correlation. Again red crosses are the mean
values and errorbars for the two groups, while the straight line is the fit to
the data. This fit has the following form:

\begin{center}
${\rm log\epsilon(He) = +0.14 \pm 0.06 \cdot [Na/Fe] + 10.95\pm 0.02}$
\end{center}

In order to verify if these stars have also a homogeneous He content,
we performed a detailed analysis of internal errors for this element. Helium was measured by
comparing the observed spectrum with 5 synthetic ones, adopting the value
that minimizes the r.m.s.\ scatter of the differences. 
The S/N of the real spectrum and the strength of the He line introduce an error in
the final He abundance which can be estimated
calculating the error on the r.m.s.\ scatter. For our data this error 
corresponds to uncertainties in the abundance of 0.06 dex.
This value must be added to the errors due to the uncertainties on atmospheric parameters.
As discussed before $\Delta$T$_{\rm eff}$=50 K gives $\Delta$log$\epsilon$(He)=0.02 dex, $\Delta$log(g)=0.06 gives
$\Delta$log$\epsilon$(He)=0.02 dex, while the error on microturbulence has no appreciable
influence.
The final uncertainty on the He abundance is given by the squared sum of all the
individual errors, and the final result is $\Delta$log$\epsilon$(He)$_{\rm tot}$=0.07 dex. 
If we compare this value to the observed dispersion (0.06$\pm$0.02 dex),
we can conclude that our HB stars are compatible with having a homogeneous He content.

\begin{figure*}[ht]
\epsscale{1.50}
\plotone{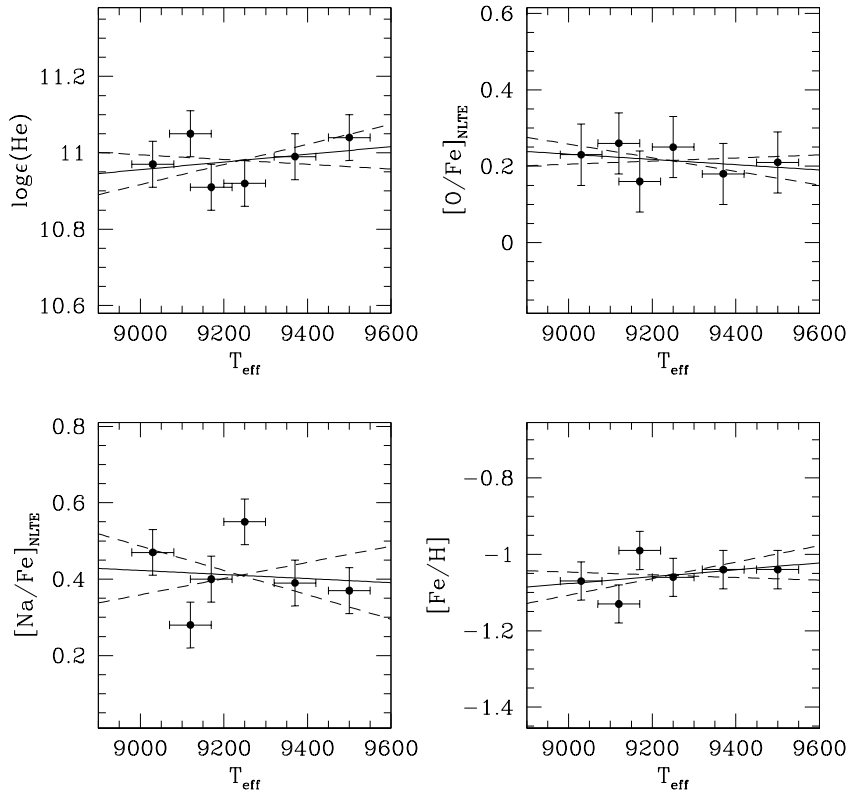}
\caption{log$\epsilon$(He), [O/Fe], [Na/Fe], and [Fe/H] vs. T$_{\rm eff}$ for
our target stars in M4. For each element we plotted the best fitting (continuous)
and the $\pm$1$\sigma$ lines (dashed)}
\label{f7}
\end{figure*}

As discussed in the introduction,
levitation and sedimentation are present in HB stars with temperatures
hotter than 11500 K. As our stars are cooler, we expect that they are not
affected by these phenomena.
With the purpose of verifying this hypothesis we plotted
log$\epsilon$(He), [El/Fe] and [Fe/H] vs. T$_{\rm eff}$ in
Fig.~\ref{f7}. For each element we plotted the best fitting (continuous)
and the $\pm$1$\sigma$ lines (dashed).
All the elements show a trend that is flat within the errors.\\
Considering also that our abundances agree well with the RGB values of \citet{Ma08}, 
we conclude that none of the elements 
measured in the present paper show evidence of levitation  
or He sedimentation.

\section{An independent photometric He estimation}

\begin{figure*}[ht]
\epsscale{1.50}
\plotone{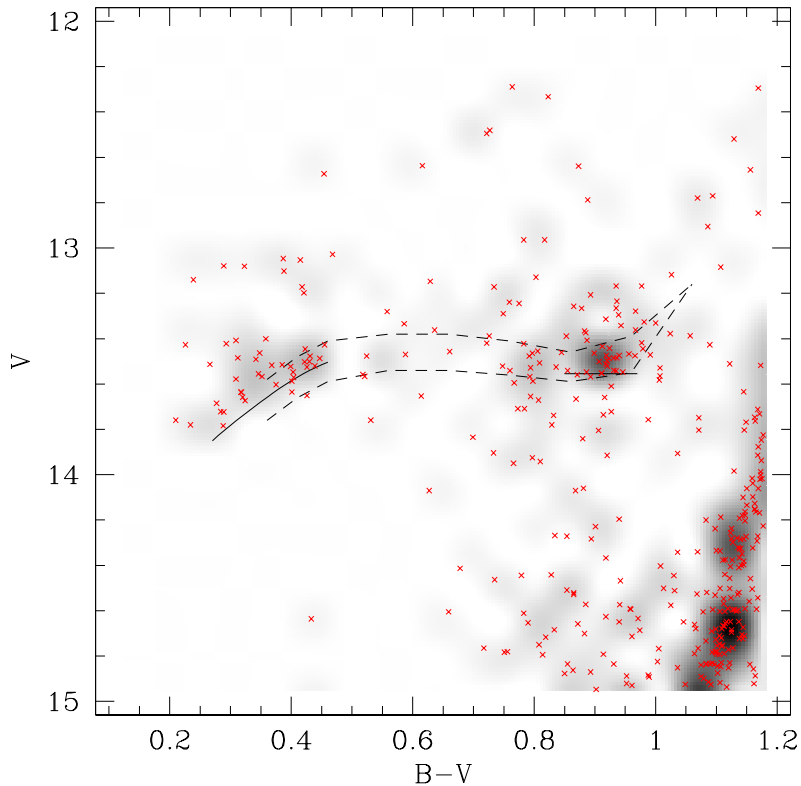}
\caption{Hess diagram of the HB. Red and blue ZAHB (solid curves) were drawn by hand. Dashed
         are the ZAHB models by \citet{Da02} with Y=0.24 (the 
         fainter) and Y=0.28 (the brighter). Single stars are shown as red crosses.}
\label{f8}
\end{figure*}

As noted by many authors (e.g. \citealt{Da02}), the brightness
of the HB depends on the He content. The higher the He content, the brighter
the HB. As our BHB stars are He enhanced, while the reddest HB stars in each cluster
are expected to be He-normal (Y$\sim$0.25, as we found in NGC 6752),
we also expect a difference in V magnitude between the two branches. 
The difference depends on the exact He difference, but for
a value of the order of $\Delta$Y$\sim$0.04 suggested by our analysis,
it should be of the order of {\bf $\sim$0.15 mag \citep{Cat09}}.
In order to verify this hypothesis we compared our photometry with
zero age HB models by \citet{Da02}. At this point the differential reddening
affecting the cluster is a problem because it blurs out the exact HB
location. To solve this issue we corrected the CMD for the differential
reddening  as explained in \citet{Sa07}. Then, in order to locate
exactly the HB, we built up the Hess diagram plotted in Fig.~\ref{f8}.

In this figure, the red HB is clearly visible as a bump at B-V=0.93, V=13.49,
while the blue HB is the overdensity covering the range 
$0.25<$B-V$<0.45$ and $13.4<$V$<13.8$.
We overplotted on both branches the HB lower envelope lines
(the continuous lines) drawn to be 3$\times\Delta$V brighter than the
faintest HB star, where $\Delta$V is the typical photometric error at the
level of the HB. Dashed lines in the plot are the zero age HB models
for the metallicity of the cluster by \citet{Da02} with Y=0.24 (fainter line) 
and Y=0.28 (brighter line). Models were shifted to the red by E(B-V)=0.36,
and  shifted in V in order to fit the lower envelope of the red HB with the model at Y=0.24.
{\it The model for Y=0.24 does not fit the lower envelope of the blue HB, 
which is brighter, as expected if it is He-enhanced}.
Then, we estimated the difference in V magnitude between the model
at Y=0.24 and {\bf the blue HB ZAHB}. It turned out to be $\sim$0.1 mag.
{\bf This implies a difference in He of the two HBs of $\Delta$Y$\sim$0.02
according to the models of \citet{Da02}, and of $\Delta$Y$\sim$0.03 according
to \citep{Cat09}.

We conclude the photometric test further supports our contention that blue HB stars in M4
are He-enhanced (Y=0.29) with respect to the red HB by $\Delta$(Y)=0.02$\div$0.03.}

\section{Comparison with literature}

Although we claim that M4 blue HB stars are He-enhanced,
several previous papers found no evidence for this.
The first is \citet{Be03}. Here the author obtained chemical abundances
(including He) for a sample of blue HB stars around the Grundahl-jump for 6
clusters: M13, M15, M3, M68, M92, NGC288. In his Fig. 22 the author reports
[He/H] value as a function of log(T$_{\rm eff}$), and apparently this does not
support our result about the He-enhancement because all stars cooler than the
Grundahl-jump are compatible with a normal He content ([He/H]$\sim$0). However
we wish to call attention to the following point. For three clusters (i.e. M13,
M15, M92) the targets cooler than the
Grundahl-jump belong to the reddest part of the HB so, as in the case of
\citet{Vi09}, they are indeed expected to be He-normal. Two of the remaining clusters (NGC288 and M68)
do not have enough points below the Grundahl-jump to derive a firm conclusion.
We are left with M3. The first impression from Fig. 22 of \citet{Be03} is that these stars
have a normal He content, but we immediately see that the errors are huge
($\sim$0.7 dex in the single measurement). Secondly, stars are located in the
intermediate region of the HB (see Fig. 1 of \citealt{Be03}), so their
He-enhancement, if any, is expected to be close to the primordial value. 
Finally we point out that all \citet{Be03} He measurements appear to be
underestimated. For example M3 has a mean He content of
[He/H]$\sim$-0.5, which translates into Y$\sim$0.10, clearly too low.
The explication is that \citet{Be03} was not interested in the absolute
abundance of He as we are, but rather in the trend of He (among the other elements)
with position along the HB. So he did not check possible systematic errors
in his methods (e.g. the He linelist). For his purposes this was not necessary,
but it makes a comparison with the present paper very problematic.

A second paper is \citet{Cat09}. Here the authors compare the HB of M3 with
theoretical models having different He content. They assert and show in their
Fig.~2 and 3 that there is no evidence of He enhancement because the model
with Y=0.25 fits well the HB. However, as \citet{Cat09} says, the fit is good,
{\it except perhaps for a small deviation of the lower envelope of the blue HB
stars in the immediate vicinity of the "knee" from the theoretical (single-Y)
ZAHB.}. The deviation is of $\sim$0.05 mag, with the stars
brighter than the model. This value is small, but clearly visible and points
toward a He-enhancement of blue HB stars in M3 with respect to red HB stars.
We would have expected an even larger deviation for the bluest HB stars of the
cluster which are expected to be even more He-rich, but in that region of the
HB the models are degenerate. We can assert that \citet{Cat09}, instead of
contradicting our finding, actually supports it.

{\bf Finally \citet{Sa08} fit the HB of NGC~1851 with their theoretical models
in order to obtain some hint about nature of the two populations of the
clusters that were identified by \citet{Mi08} as a bimodal SGB. They can fit the HB with
two models. In the first the two populations are assumed to have an age
difference of 1 Gyrs and the same (primordial) He, CNO, anf Fe content. In the
second they are coeval and have different CNO (but the same primordial
He and the same Fe). Apparently there is no room for an He-enhancement. First
of all we notice that more recent papers \citep{Vi10,Ca10b} found that stars
in NGC~1851 have the same CNO content but different Fe ($\sim$0.06 dex). A
difference in Fe have an impact on the age difference that is $\sim$0.5 Gyrs
(assuming the same CNO).
So none of the models by \citet{Sa08} is appropriate to fit the HB. On the
other hand if we consider in their Fig. 2 (e.g. lower panel) the line
that connect the red part with coolest part of the blue synthethic HB and
project it on the observed HB (in order to fit the observed red HB), we can
see that the observed blue HB is slightly brighter than this line. 
According to any HB model that assume the same CNO content (including
\citealt{Sa08}, see their Fig. 1) this is an indication that the blue HB is
He-enhancend with respect the red.
This fact is confirmed by a recent paper \citep{Gr12}. Here the authors
obtained a spectroscopic estimation of
Y=+0.29$\pm$0.05 for the BHB. The error is quite large, but they show
photometrically that the HB can be fitted only assuming Y=+0.248 and Y=+0.280
for the red and blue HBs respectivelly.

We conclude this section by noting that literature statements against the
presence of He-rich stars in GCs are not conclusive, and that they are refuted
by a new interpretation of the data or by new results.
}

\section{Discussion}

Correlations and anticorrelations of chemical elements observed in GCs (i.e. Na vs. O and Mg vs. Al)
are attributed to contamination by products of the H-burning process at
high temperature (\citealt{La93}, \citealt{Pr07}), when N is produced at the
expense of O and C, and proton capture on Ne and Mg produces Na and Al (CNO,
{\bf NeNa, and MgAl} cycles).
\citet{Gr01} demonstrated that this contamination is present also at the level
of the MS. This means that it is
not the result of a mixing mechanism present when a star leaves the MS,
but it is rather due to primordial pollution of the interstellar material from
which stars were formed. The mixing mechanism along the RGB can have an
effect, but only as far as C and N are concerned \citep{Gr00}.

Pollution must come from more massive stars. GCs must have experienced some 
chemical evolution at the beginning of their life (see \citealt{Gr04} for extensive references).
The main product of H-burning is He and a He enhancement is then expected to be
present in stars with an enhancement of N, Na, and Al. The main classes of
candidate polluters are: fast-rotating massive main-sequence stars \citep{De07},
intermediate-mass AGB stars \citep{Da02}, and also massive binary stars \citep{Mi09}.
All these channels can potentially
pollute the existing interstellar material with products of complete CNO
burning, including He (see \citealt{Re08} for an extensive review).

Recently \citet{Vi11} showed that for M4 the best candidates that can
reproduce the abundance pattern observed for RGB stars are massive main-sequence stars.

In the pollution scenario, a first generation of O-rich and Na-poor stars
(relative to a second stellar generation)
is formed from primordial homogeneous material, which must have been
polluted by previous SN explosions. 
This generation also has a primordial or close to primordial He content (Y=0.24$\div$0.25).
Then the most massive stars of this pristine stellar generation
pollute the interstellar material with products
of the CNO cycle. This material is kept in the cluster due to the relatively 
strong gravitational field
\citep{De08} , and it gives rise to 
a new generation of O-poor and Na-rich stars.
This population should also have been He-enhanced (Y=0.27-0.35, depending on
the major polluter and on the amount of retained material, \citealt{Da02,Bu07}).
Also the abundance of other elements (including s-process elements) 
may differ in stars of the first or second generation.

In the MS phase, He-rich stars evolve more rapidly than He-poor ones,
so He/Na-rich (and O-poor) stars presently at the turn-off or in later phases of 
evolution should be less massive than He/Na-poor (and O-rich) ones.
In this framework, \citet{Da02} and \citet{Ca07} proposed that a 
spread in He, combined with mass loss along the RGB, 
may be the main ingredient to naturally reproduce the whole HB
morphology in GCs, as discussed in the introduction. 
According to this scenario, primordial He-O rich and
Na poor stars end up on the redder part of the HB, while stars with extreme abundance 
alterations (strong He enhancement, O-poor and Na-rich) may end up on the
bluer part of the HB, if they have experienced enough mass loss during the RGB
phase.
The extensive comparison between the distribution of colours along
the HB and the Na-O anti-correlation by \citet{Gr10} also supports
this view.
Also age must play a role, because it determines the mean mass of a
star that reaches the HB \citep{Gr10,Do10}.

We stress that the HB represents the ideal 
locus to investigate the effects of chemical anomalies in GC stars, as 
the HB is a sort of amplifier of the physical conditions
that are the consequence of the star composition and previous evolution.

In this scenario, in M4, we expect that the progeny of RGB 
He-normal, O-rich and Na-poor stars should reside in the
red part of the HB. Therefore, it is not surprising that
the observed red HB stars of \citet{Ma11} 
are all O-rich, Na-poor. But they are too cold to have 
their He content measured directly.
\citet{Ma11} show also that blue HB stars are all O-poor and Na-rich.
Here we reinforce this result with higher S/N spectra and more precise
measurements. But we go further, measuring the He content of the blue HB. 
All our stars turned out to be He-enhanced {\bf (Y=0.29)}.
By comparing photometry with HB models, we {\bf showed that the blue HB is brighter
than the red, as expected if they have a different He content}.

If we consider both the He  and Na/O content of our targets, our result
strongly confirms the hypothesis suggested by  \citet{Da02}, \citet{Ca07},
\citet{Gr10}, and shows that He (coupled with light-element spread) is one of
the best candidates (together with the metal content and age) to explain the
HB morphology of GCs and thus is a strong candidate for the 2$^{nd}$ parameter
(or for the 3$^{rd}$ according to \citealt{Gr10}).

\section{Summary}

We studied a sample of BHB stars in M4 with a temperature in the range 9000-9500 K,
with the aim of measuring their He and Na/O content.
Targets were selected in order to be hot enough to show the He feature at 5875 \AA,
but cold enough to avoid the problem of He sedimentation and metal levitation
affecting HB stars hotter than 11500 K.
Thanks to the high resolution and high S/N of our spectra, we were able to measure
He abundances for all our stars.
This is only the second, direct measurement of He content from high
resolution spectra of GC stars in this T$_{\rm eff}$ regime with the aim
to test the current models of GC formation and multiple-populations.

Our sample of stars turns out to have [Fe/H]=-1.06$\pm$0.02,
value that agrees well with the literature, and belong 
to the O-poor/Na-rich population of M4 found in the RGB region by
\citet{Ma08}.

Our targets have a mean He content of Y=0.29\\$\pm$0.01 (internal error) which is significantly
larger than the value found for redder BHB stars in NGC~6752 (Y=0.25$\pm$0.01),
using the same observational set up and data reduction procedures.
Our best estimation for the systematic error is $\Delta$Y=0.01 (or
0.014). However it does not affect the comparison with NGC~6752 or the
photometric estimation of the red HB He content.
{We compared also the brightness of the red and the blue HB of the cluster,
finding that the latter is $\sim$0.1 mag brighter then the former.
This result is quite strong, and we can estimate an enhancement in the He content for the
stars in the blue vs. red HB of $\Delta$Y=0.02$\div$0.03.}
This is what we expect if stellar position on the HB is driven by its He and
metal content. {\bf Our combined evidence strongly suggests that stars within the same globular
cluster and among different globular clusters have different He content}, and
that the O-poor/Na-rich population is also He-enhanced {\bf with respect the
O-rich/Na-poor population}, as suggested by many theoretical studies. He is
thus a leading contender for the 2$^{nd}$ parameter.

Our results are consistent with theoretical studies which predict that, for a
given metallicity and age, the position of a star in the HB is driven by its He
content, and that the spread of stars along the HB must be related to the Na/O spread
visible at the level of the RGB.

\acknowledgments

S.V. and D.G.gratefully acknowledge support from the Chilean
{\sl Centro de Astrof\'\i sica} FONDAP No. 15010003
and the Chilean Centro de Excelencia en Astrof\'\i sica
y Tecnolog\'\i as Afines (CATA).\\
G.P. and R.G. acknowledge support by MIUR under
the program PRIN2007 (prot.\ 20075TP5K9), and by INAF
under PRIN2009 'Formation and Early Evolution of Massive
Star Clusters'.\\
The authors gratefully acknowledge also the referee that helped clarify
a number of important points.

\end{document}